# Network Traffic Anomalies Detection and Identification with Flow Monitoring

Huy Anh Nguyen, Tam Van Nguyen, Dong Il Kim, Deokjai Choi
Department of Computer Engineering, Chonnam National University, Korea
Email: anhhuy@gmail.com, vantam@gmail.com, hangaram@chonnam.ac.kr, dchoi@chonnam.ac.kr

*Abstract*-Network management and security is currently one of the most vibrant research areas, among which, research on detecting and identifying anomalies has attracted a lot of interest. Researchers are still struggling to find an effective and lightweight method for anomaly detection purpose. In this paper, we propose a simple, robust method that detects network anomalous traffic data based on flow monitoring. Our method works based on monitoring the four predefined metrics that capture the flow statistics of the network. In order to prove the power of the new method, we did build an application that detects network anomalies using our method. And the result of the experiments proves that by using the four simple metrics from the flow data, we do not only effectively detect but can also identify the network traffic anomalies.

## I. INTRODUCTION

Internet traffic measurement is essential for monitoring trends, network planning and anomaly traffic detection. In general, simple packet- or byte-counting methods with SNMP have been widely used for easy and useful network administration. In addition, the passive traffic measurement approach that collects and analyzes packets at routers or dedicated machines is also popular. However, traffic measurement will be more difficult in the next-generation Internet with the features of high-speed links or new protocols such as IPv6 or MIPv6.

Traffic measurement at high speed links is challenging because of fast packet-processing requirement. Though packet-level measurement can describe the detailed traffic characteristics, it is not easy to support high-speed line rates of multi-gigabit per second. Moreover, standalone systems for packet-level traffic monitoring will be expensive for the wide deployment and easy management in a large-scale network. Hence, ISPs or big ASes will generally prefer the flow-level traffic measurement approach that could be easily embedded into routers or switches to dedicated packet-level traffic monitoring systems. Currently, flow-level measurement modules at routers such as Cisco NetFlow [1] have become popular, because flow-level measurement could generate useful traffic statistics with a significantly small amount of measured data.

While monitoring the traffic and detecting anomalous activities is important, it is equally important to keep the rate of *false alarms* low. A high false alarm means that the genuine events will be lost in the "snow" of false events. Suppose that we apply one's statistical anomaly detection method on large networks (with thousands of switches and routers involved and millions of users), even a very small false alarm rate may result in enough false alarms to overwhelm that network operation staff. In the worst case, false alarms undermine anomaly detection, as operation staff tire of reacting to false alarms, and ignore or turn the system off entirely. Currently, researchers are still struggling for a simple but robust method for anomaly detection, with high detection rate and low false alarm.

Although anomaly detection has been addressed in many prior projects, there is the fact that few works have been succeeded in statistically characterized different types of network traffic flow anomalies. Furthermore, most anomaly detection methods are limited to analyzing the entire traffic as one entity, which makes them unable to quantify network anomalies, and their validities are affected when many anomalous activities occur simultaneously. From that we see the need for a method that can effectively detect and classify network anomalies based on flow statistics.

In this paper, we analyze traffic flow information to detect abnormal behaviors. Traffic flows are created from all packets captured from a network link. There are many network anomalies, but whenever a network anomaly occurs, traffic behavior will change abruptly. These changes can be inspected by tracking various parameters of traffic flows. Not only monitoring fundamental flow parameters such as flow size, number of packets, which represent major flow features, we also statistically collect certain extended metrics in order to infer the connection trends of flows. These flow metrics are not only used for detecting but also for identifying a number of network anomalies as flows in various anomalies has their distinctive set of flow metric values. Within each metric, we use a standard technique called Holt-Winters [6] to extract the anomaly indicators. Holt-Winters algorithm will constantly monitor each single metric, and maintain a list of historical data for the anomaly detection purpose. Whenever the value of a monitored metric goes outside the range of predicted data, Holt-Winters algorithm will raise an anomaly flag on the metric. The network is then likely to be undergoing anomalous activities; this is the case when the system needs attention from the network operation staff.

Experience has shown that, just giving trivial alerts does not turn out to be very helpful for the problem diagnosing. Usually, alerts just give limited amount of clue for the operation staff to identify and react to the problem. And in some cases, a false alarm may cause a huge waste in the effort when network operators try to seek for a problem that does not exist. Our method alleviates the issue. When flags are raised indicating anomalous activities, the network administrator will then use the combination of the flag to identify the type of network anomaly. Giving out the clue for the anomaly will provide great information for network administrators to trace back and perform network forensic.

The organization of this paper is as follows: Section 2 provides the overview of the related work. Then, in Section 3, we describe the core of our anomaly detection method, the Holt-Winters algorithm. Section 4 will describe the most popular malicious network traffic attacks and the way to identify each ones, the overall of our technique will be summarized at the end of Section 4. Section 5 describes how to build the flow monitoring system, the flow monitoring application that uses our new method and its experimental results. Finally, the conclusions and future directions will be given in Section 6.

## II. RELATED WORK

As stated above, anomaly detection has been addressed in many prior projects, and previous works have primarily focused on security tasks (detecting DDoS attacks, worms, or other intrusion…). In many cases, providers use very simple techniques for anomaly detection, such as fixed threshold, packet capturing and analyzing… for the case of DDoS detection, Cisco and Juniper [2] also embedded in their routers a simple flood attack protection based on threshold technique. For the 3[rd] party solutions in network traffic anomaly detection such as D-Ward, Multops… they also tried to define thresholds for TCP, UDP and ICMP applications, then an attack or anomalous activities will be detected and given alarm whenever a threshold is exceeded. These methods are quite limited since there is no fixed threshold for different kind of networks. Also these methods call for experienced network operators to define thresholds and constantly monitor and modify them.

There has however been some more sophisticated works in the detection and analysis of network anomalies. Instances are [7] [8] [9] [10] [11]. Of these, the most directly relevant to this paper is [11] which tests the use of Holt-Winters forecasting technique for network anomaly detection. In this paper, we also focus on exploiting Holt-Winters algorithm for the anomaly detection purpose, but with a different way. Holt-Winters will be used for monitoring the 4 predefined metrics and the network administrator will identify the anomaly when it happens based on anomaly flags of the metrics. (Of course, we did make some modifications on the original Holt-Winters algorithm in our experimental implementation.) Although the method we use here is very simple, it gives good performance and detecting result. And it's quite open in the mean that other researchers can easily apply our techniques in the combination with theirs to improve their detection schemes.

## III. HOLT-WINTERS FORECASTING TECHNIQUE

Holt-Winters Forecasting is a sophisticated algorithm that builds upon exponential smoothing. Holt-Winters Forecasting rests on the premise that the observed time series can be decomposed into three components: a baseline, a linear trend, and a seasonal effect. The algorithm presumes each of these components evolves over time and this is accomplished by applying exponential smoothing to incrementally update the components. The prediction is the sum of the three components:

$$y_{t+1} = a_t + b_t + c_{t+1-m}. \quad (1)$$

The update formulas for the three components, or coefficients $a, b, c$ are:

- Baseline ("intercept"):
$$a_t = \alpha\,(\,y_t + c_{t-m}\,) + (\,1 - \alpha\,)(\,a_{t-1} + b_{t-1}\,)\,. \quad (2)$$
- Linear Trend ("slope"):
$$b_t = \beta\,(\,a_t - a_{t-1}\,) + (\,1 - \beta\,)\,b_{t-1}. \quad (3)$$
- Seasonal Trend:
$$c_t = \gamma\,(\,y_t - a_t\,) + (\,1 - \gamma\,)\,c_{t-m}. \quad (4)$$

As in exponential smoothing, the updated coefficient is an average of the prediction and an estimate obtained solely from the observed value $y_t$, with fractions determined by a model parameter ($\alpha, \beta, \gamma$). Recall $m$ is the period of the seasonal cycle; so the seasonal coefficient at time $t$ references the last computed coefficient for the same time point in the seasonal cycle.

The new estimate of the baseline is the observed value adjusted by the best available estimate of the seasonal coefficient ($c_{t-m}$). As the updated baseline needs to account for change due to the linear trend, the predicted slope is added to the baseline coefficient. The new estimate of the slope is simply the difference between the old and the new baseline (as the time interval between observations is fixed, it is not relevant). The new estimate of the seasonal component is the difference between the observed value and the corresponding baseline.

$\alpha, \beta$ and $\gamma$ are the adaptation parameters of the algorithm and $0 < \alpha, \beta, \gamma < 1$. Larger values mean the algorithm adapts faster and predictions reflect recent observations in the time series; smaller values means the algorithm adapts slower, placing more weight on the past history of the time series. These values should be optimized when the algorithm is implemented.

## IV. NETWORK TRAFFIC ANOMALIES

Within the scope of this paper, we address a set of network-centric anomalies which exhibit abnormal changes in network traffic. Other kinds of application-related attacks such as buffer overflow, guess password … are not included in the proposed method because their malicious purposes are carried within the payload of the packets.

*A. UDP flood*

A UDP flood attack is a kind of DoS attack. An attack can be initiated by sending a large number of UDP packets to random ports on a remote host. As the result, the distant host will check for the application listening on this port. After seeing that no application listens on the port, the host will reply with an ICMP "Destination Unreachable" packet. Thus, for a large number of UDP packets, the victimized system will be forced into sending many ICMP packets, eventually leading it to unreachable by other clients. If enough UDP packets are delivered to the ports on the victim, the system will go down.

When a network undergoes a UDP flood attack, it is expected from the perspective view of flow that there will be a tremendous amount of UDP datagrams from one or many outside sources to a specific destination on the network. In order to detect and identify this type of attack, we need to work with the *flow size* and *packet count* of the flows. Flows of UDP flood contain many large-sized packets with the intention to overwhelm the destination. Therefore, they will likely to have large values of these two metrics. We then hereby define the first two metrics:

- **TotalBytes**: total volume of flows in bytes.
- **TotalPackets**: total packets in flows.

The monitoring process over the first two metrics may plot a noticeable increase of the metrics when network under attack.

*B. ICMP flood*

ICMP flood or ping flood is a simple attack where the attacker overwhelms the victim with ICMP Echo Request (ping) packets with different sizes. This method is an upgrade from its predecessor, the Ping-of-Death attack. PoD tries to send an over-sized ping packet to the destination with the hope to bring down the destination system due to the system's lack of ability to handle huge ping packets. Ping flood brings the attack to a new level by simply flood the victim with huge ping traffic. The attacker hopes that the victim will too busy responding to the ICMP Echo Reply packets, thus consuming outgoing bandwidth as well as incoming server bandwidth.

Similar to UDP flood, ICMP attack also generate a huge amount of data towards the destination. So using TotalBytes and TotalPackets is enough to plot this type of attack. Of course, using the same method means that we can't distinguish ICMP from UDP flood since we use the same signature for the two types of attack. This problem can easily be solved by using another metric for monitoring the total number of ICMP or UDP traffic going into the network; then when the attack occurs, we can easily distinguish between the two. In this paper, however, we will treat these two attacks the same way.

*C. TCP SYN attack*

This method takes advantage of a flaw in how many hosts implement the TCP three-way handshake. When host B receives the SYN request from host A, it must keep track of the partially opened connections in a "listening queue" for at least *n* seconds (e.g.: 75 seconds). Many host implementations can only keep track of a very limited number of connections. A malicious host can exploit the small size of the listen queue by sending multiple SYN requests to a host, but never reply to the sent back SYN and ACK. By doing so, the destination host's listening queue will be quickly filled up, and it will stop accepting new connections.

Due to the characteristic of TCP SYN, the effect of this attack on network traffic is quite different from the two previous attacks. Flows of TCP SYN have small values. The real number of packets differs from each attack implementations, but usually the flows contain less that 3 small size packets. Since the attacker just has to generate a small amount of network traffic in order to bring down a particular host, we can't rely on TotalBytes or TotalPackets to plot this kind of attack; we need to define a new metric that can group together all flows in the attack:

- **DSocket**: number of flows with similar volume to the same destination socket (IP and port).

*D. Portscan*

A portscan attack is carried out with a port scanner, a piece of software to search a network host for open ports. A port scanner is often used by network administrators to check the security of their networks, and it also used by hackers to compromise the system security. Many exploits rely upon port scans, for example to find open ports and send large quantities of data in an attempt to trigger a condition known as *buffer overflow*, or to send some specific port data packets with malicious purposes …

A portscan operation will result a big number of packets sent from a remote host to a destination on the network, but with different destination ports. Flows in portscan are small flows with the size of only several bytes and packet count of 2 or 3. This malicious activity cannot be detected with the three metrics we already have. In order to gather together all flows in a portscan attack for the detection purpose, we need to define another metric that has the capability to aggregate all these flows:

- **DPort**: number of flows that have a similar volume, same source and destination address, but to different ports.

*E. Anomaly detection and identification*

TABLE I
ANOMALY IDENTIFICATION TABLE WITH THE FOUR METRICS

| Anomaly | *TotalBytes* | *TotalPackets* | *DSocket* | *DPort* |
|---|---|---|---|---|
| Flooding | High | High | Normal | Normal |
| TCP SYN | Normal | Normal | High | Normal |
| Portscan | Normal | Normal | Normal | High |
| Normal | Normal | Normal | Normal | Normal |
| Other | - | - | - | - |

Bases on the analysis above, we can summarize our method within the flow distribution table for the flow metrics as shown in Table I. In our Holt-Winters implementation, the deviation or the time-series data at the time $t$, $d_t$, and the normal traffic range are calculated by:

$$d_t = \gamma | y_t - \hat{y}_t | + (1-\gamma) d_{t-m}. \quad (5)$$

And:

$$\hat{y}_t \pm \delta \times d_{t-m}. \quad (6)$$

If $y_t$ is above or below the range above, then the traffic is HIGH or LOW, respectively. Whenever a LOW traffic anomaly occurs, is most likely to be network device outages. That type of anomalies is not the main addressing point in our research; therefore we will focus on the HIGH traffic anomaly. Based on the anomaly flags raised by the Holt-Winters algorithm, the network administrator can spot the anomaly and quickly identify it.

Our proposed method has some advantages over the others. Our method give anomaly detection based on the combination of different flow metrics. The distribution table of our method is simple, despite its good performance. Whenever an anomaly is detected, the combination of the flags gives us the idea about what is happening in the network. Experience has shown that this information is at great value to network operators to perform trace back and problem forensic. Besides, we do not treat the network traffic as one entity. The detection scheme is based on the four metrics. Therefore, we can still distinguish many types of anomalies when they occur simultaneously (e.g.: if network under both TCP SYN and Portscan attack, DSocket and DPort will both raise the alarm).

## V. EXPERIMENTS

### A. Flow data collecting

In order to conduct the experiment of our research, we first start with defining the data structure to be gathered. The data have to be simple, but contains enough information for the calculation of the four flow metrics. There are many flow data structures that can be used to acquire the objective above, examples are sFlow, nFlow, NetFlow v5, v9 [1], IPFIX [3]. Cisco NetFlow v5 is the most commonly used data structure for flow-level monitoring in current IPv4 networks. However, NetFlow v5 has a fixed key structure and is not capable of monitoring various kinds of new network protocols and IPv6 traffic [13]. To tackle the drawback of NetFlow v5, IPFIX is equipped with the flexible and extensible template architecture, and it uses reliable SCTP/TCP as the default transport protocol. Therefore in our research, we will collect the data in IPFIX format with the IPFIX template depicted in Fig. 2.

However, there is one problem working with IPFIX. Since IPFIX is a new standard and not yet available in all network devices; our router, unfortunately, is incapable of collecting flow data in IPFIX format. To fix the problem, we use a modified version of JNCA [4] to collect flows in NetFlow v9 format at the flow collector, and then we convert the relevant fields into IPFIX and insert the data into a MySQL database for further analysis.

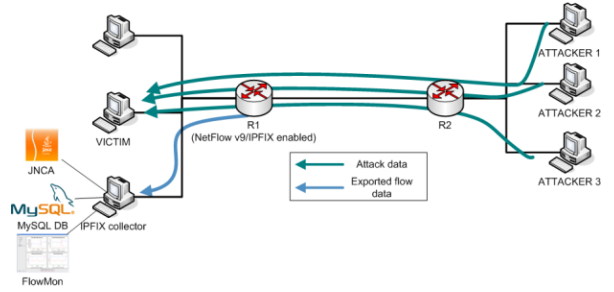

Fig. 3. The experimental environment for measurement of flow parameters.

### B. Anomaly network testbed

An IPFIX network testbed for the experiments is depicted in Fig. 3. We tried to shape the real attack environment. There are two networks: attacking network and destination network. They are connected together by two or more routers. In order to simulate "beautiful" flow data, we made the victim (or destination of attacks) a web server with common services to provide network traffic including: web browsing, email, database access, file transfer ... All accesses to these services are legitimate traffic.

Router R1 is the gateway router of the destination network, it has to be able to export flow data in NetFlow v9/IPFIX format. The exported flow data will be sent to the IPFIX collector, which located in the destination network. On the IPFIX collector, we run some essential programs to collect and analyze flow data: JNCA for converting NetFlow v9 to IPFIX format, MySQL database for storing flow data and FlowMon (the application built by us) for monitoring flow data and raising anomaly flags. The malicious data can be carried out by The Hacker Toolkit [5], or many attacking tools such as stream2, synhose, synk7, synend, hping [14] ... however, the usage of these malicious tools will not be stated in our paper.

| Version = 10 | Msg. Length | |
|---|---|---|
| Export Timestamp = 2007-03-02 23:59:00 | | IPFIX Header |
| Sequence Number = 0 | | |
| Source ID = 12345678 | | |
| Set ID = 2 (Template) | Set Length | |
| Template ID = 256 | Field Count = 11 | |
| ProtocolID = 4 | Field Length = 1 | |
| SrcIPv6/v4Addr = 27/8 | Field Length = 16/4 | |
| DestIPv6/v4Addr = 28/12 | Field Length = 16/4 | |
| L4SrcPort = 7 | Field Length = 4 | IPFIX Template Set |
| L4DstPort = 11 | Field Length = 4 | |
| NextHeader = 193 | Field Length = 4 | |
| FlowLabel = 31 | Field Length = 4 | |
| FirstTime = 22 | Field Length = 4 | |
| LastTime = 21 | Field Length = 4 | |
| OctetDeltaCount = 1 | Field Length = 4 | |
| PacketDeltaCount = 2 | Field Length = 4 | |
| Record 1, Field 1 | | |
| . | | |
| Record 1, Field 10 | | IPFIX Data Set |
| Record 2, Field 1 | | |
| . | | |
| Record 2, Field 10 | | |

Fig. 2. IPFIX template with the highlighted fields involved in the monitoring metrics

## C. Experimental results

TABLE II
DETECTION RESULT FOR SIMULATION DATA

| Anomaly | Inserted / Detected |
|---|---|
| Flooding | 28 / 26 |
| Distributed flooding | 11 / 10 |
| TCP SYN | 13 / 13 |
| Portscan | 15 / 15 |
| Other anomalies | 0 / 0 |

At first, we set the testbed network up and running for 2 days without any anomalies to model the normal traffic pattern of the network. Holt-Winters algorithm will constantly monitor and study the normal data. This will provide the baseline values of normal network condition. Then we start to insert the anomalous data for 3 hours with a set of network anomalies including: UDP flood, ICMP flood, TCP SYN and Portscan. Attack data will be destined to the victim. The attack scenario is staged by the tools talked above and each attack will occurs in about 5 minutes. Flow data from attackers will be captured by the router and exported to the exporter for storing and analyzing. The detection results are shown in Table II.

In Table II, flooding attacks include both UDP and ICMP flooding. Just as we expected, almost all the inserted attacks are detected by FlowMon right after they occurred. We tested the validity of our method in the scenario of multiple anomalies occur at the same time and with different anomalies combination: UDP + TCP, ICMP + Portscan … FlowMon still be able to raise the exact flags according to the anomaly type. In the case of TCP SYN and Portscan, all the small malicious flows that go through the router were plotted by DSocket and DPort, and all these attacks were detected and identified. Our method still works well when we switch the attacking method from 1 source – 1 destination (flooding) to multi sources – 1 destination (Distributed flooding).

However, the validity of our method is affected at the end of our 3 hours experiment. After more than 2 hours of gradually inserting the anomalies, we continuously tried to insert flooding/distributed flooding attacks into the destination network in high density. Here then FlowMon could not detect one of the last attacks. The phenomenon can be explained by examining Holt-Winters algorithm. One of the features of Holt-Winters algorithm is to gradually adapt to the current values of the time series. When anomalies happen constantly, after a certain amount of time, the anomaly data became familiar. Holt-Winters algorithm will adapt itself to the anomaly data, and consider anomaly data normal. To prevent that issue from happening, network operators should quickly react to the anomalies, and have solutions to stop them from occurring for a long time. How to protect the network from anomalies, however, is outside the scope of this paper.

## VI. CONCLUSION

So far in our research, we proposed a new method for network traffic anomaly detection with four predefined metrics: *TotalBytes*, *TotalPackets*, *DSocket*, *DPort*. We also built a network testbed and a program called *FlowMon* to certify the feasibility of our method. Based on the algorithm and experiment verification, the method has proved its efficiency in anomaly detection. This research was dedicated to deal with network centric anomalies that exhibit abnormal changes in network traffic, and our contribution is a new lightweight method that does not only quickly detect network anomalies, but can also pin point what kind of anomaly it is. The method is simple, but it should be seen as an advantage, as simplicity makes the method scalable, and more easily extendible to include other features, for instance, new metrics for other kinds of network anomaly, or applying wavelet analysis for better anomaly detection rate. That's why the method has great potential to be re-used in further researches.

For future direction from this research, we would like improve the detection result by implementing more sophisticated algorithms for monitoring the metrics, such as ARAR algorithm [6] or wavelet analysis technique, then we can choose the best algorithm for each case. Besides, our method is open in the mean that whenever an unknown anomaly occurs, we can grasp its abnormal behavior and flow characteristics for further study.

ACKNOWLEDGMENT

This research was supported by Korea Telecom Inc.